\documentclass[doublecol]{epl2}

\title{Binary birth-death dynamics and the expansion of cooperation by means of self-organized growth}
\shorttitle{Binary birth-death dynamics and the expansion of cooperation by means of self-organized growth}
\author{Attila Szolnoki,\inst{1,2} Alberto Antonioni, \inst{3} Marco Tomassini,\inst{3} Matja{\v z} Perc\inst{2}}
\shortauthor{Szolnoki et al.}

\institute{\inst{1}Institute of Technical Physics and Materials Science, Research Centre for Natural Sciences, Hungarian Academy of Sciences, P.O. Box 49, H-1525 Budapest, Hungary\\
\inst{2}Faculty of Natural Sciences and Mathematics, University of Maribor, Koro{\v s}ka  cesta 160, SI-2000 Maribor, Slovenia\\
\inst{3}Information Systems Institute, HEC, University of Lausanne, CH-1015 Lausanne, Switzerland}


\abstract{Natural selection favors the more successful individuals. This is the elementary premise that pervades common models of evolution. Under extreme conditions, however, the process may no longer be probabilistic. Those that meet certain conditions survive and may reproduce while others perish. By introducing the corresponding binary birth-death dynamics to spatial evolutionary games, we observe solutions that are fundamentally different from those reported previously based on imitation dynamics. Social dilemmas transform to collective enterprises, where the availability of free expansion ranges and limited exploitation possibilities dictates self-organized growth. Strategies that dominate are those that are collectively most apt in meeting the survival threshold, rather than those who succeed in exploiting others for unfair benefits. Revisiting Darwinian principles with the focus on survival rather than imitation thus reveals the most counterintuitive ways of reconciling cooperation with competition.}

\begin{document}

\maketitle

In \textit{The Origin of Species}, Darwin laid out a beautiful theory according to which generations of organisms change gradually over time to give rise to the astonishing diversity of life we witness today. Only organisms that survive long enough to reproduce are able to pass on their genetic material, and in time those characteristics that allow survival and reproduction become more common \cite{dawkins_76}. Individual fitness is key to success, as those who fail to reproduce are destined to disappear through natural selection. But if only the fittest survive, why is there so much cooperation in nature? Eusocial insects like ants and bees are famous for their large-scale cooperative behavior, even giving up their own reproductive potential to support that of the queen \cite{wilson_71}. Cooperative breeding in birds prompts allomaternal behavior where helpers take care for the offspring of others \cite{skutch_co61}. Microorganisms cooperate with each other by sharing resources and joining together to form biofilms \cite{nadell_fems09}. Humans have recently been dubbed supercooperators \cite{nowak_11} for our unparalleled other-regarding abilities and cooperative drive. But why should an organism carry out an altruistic act that is costly to perform, but benefits another? Altruistic cooperation is the most important challenge to Darwin's theory, and it is fundamental for the understanding of the main evolutionary transitions that led from single-cell organisms to complex animal and human societies \cite{maynard_95}.

\begin{figure*}
\begin{center}
\includegraphics[width = 14.4cm]{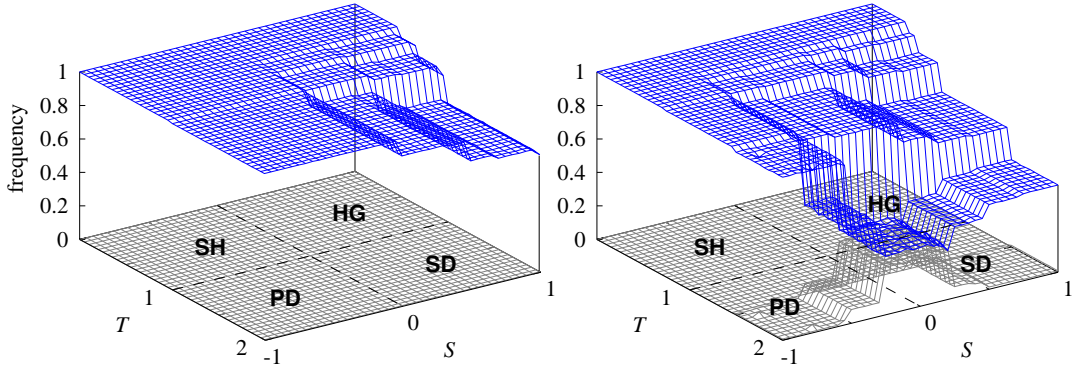}
\caption{\label{wire}Stationary frequency of cooperators (blue wire) and empty sites (gray wire) on the $T-S$ parameter plane for $Q=2$ (left) and $Q=1.5$ (right). The square lattice ($L=400$) with nearest-neighbor interaction range ($R_1$) was used, and initially cooperators, defectors and empty sites were distributed uniformly at random in the whole lattice. Dashed lines delineate the four different types of social games.}
\end{center}
\end{figure*}

Hamilton's kin selection theory has been applied prolifically to solve the puzzle \cite{hamilton_wd_jtb64}, resting on the fact that by helping a close relative to reproduce still allows indirect passing of the genes to the next generation. But since cooperation is common not only between relatives, other mechanisms have also been identified, including various forms of reciprocity and group selection \cite{nowak_s06}. Network reciprocity in particular \cite{nowak_n92b}, has recently attracted considerable attention in the physics community, as it became clear that methods of nonequilibrium statistical physics \cite{binder_88, castellano_rmp09, furtano_pr10, marsili_pa99} can inform relevantly on the outcome of evolutionary games on structured populations \cite{szabo_pr07, roca_plr09, perc_bs10, brede_pone13, wang_z_pre12, du_csf13, tanimoto_pre12}. While the basic idea behind network reciprocity is simple --- cooperators do better if they are surrounded by other cooperators --- the manifestation of this fact and the phase transitions leading to it depend sensitively on the structure of the interaction network and the type of interactions \cite{szabo_pr07}, as well as on the number and type of competing strategies \cite{szolnoki_prl12}. Physics-inspired studies have led to significant advances in our understanding of the evolution of cooperation, for example by revealing the importance of time scales in evolutionary dynamics \cite{roca_prl06}, the positive impact of heterogeneity of interaction networks \cite{santos_prl05}, the dynamical organization of cooperation \cite{gomez-gardenes_prl07} in conjunction with population growth \cite{poncela_njp09, melbinger_prl10}, as well as the emergence of hierarchy among competing individuals \cite{lee_s_prl11}.

While the infusion of physics is a relatively recent development, evolutionary game theory \cite{weibull_95, hofbauer_98, nowak_06} is long established as the theory of choice for studying the evolution of cooperation among selfish individuals. Competing strategies vie for survival and reproduction through the maximization of their utilities, which are traditionally assumed to be payoffs that are determined by the definition of the contested game. The most common assumption underlying the evolution on structured populations has been that the more successful strategies are imitated and thus spread based on their success in accruing the highest payoffs \cite{szabo_pr07, roca_plr09, perc_bs10}. As such, imitation has been considered as the main driving force of evolution, reflecting the individual struggles for success and the pressure of natural selection. In harsh environments, like bacterial colonies during growth \cite{maharjan_el13}, however, imitation may be prohibitively slow, and the extreme conditions may render a strategy viable or not in a binary way. Either the conditions for survival by an individual are met and may lead to offspring, or they are not and the individual perishes. Here we explore the consequences of such a binary birth-death evolutionary dynamics on structured populations, which effectively shifts the focus of Darwinian selection from the imitation of the fittest to the survival of the viable.

We study pairwise evolutionary games on a square lattice of size $L^2$, where mutual cooperation yields the reward $R$, mutual defection leads to punishment $P$, and the mixed choice gives the cooperator the sucker's payoff $S$ and the defector the temptation $T$. Within this setup we have the prisoner's dilemma (PD) game if $T>R>P>S$, the snowdrift game (SD) if $T>R>S>P$, and the stag-hunt (SH) game if $R>T>P>S$, thus covering all three major social dilemma types. Without loss of generality we set $R=1, P=0, 0 \le T \le 2$, and $-1 \le S \le 1$, as illustrated in Fig.~\ref{wire}. We note that $T<1$ and $S>0$ quadrant marks the harmony game (HG), which however does not constitute a social dilemma. Initially, either the whole or part of the lattice is populated by cooperators ($C$) and defectors ($D$), who are distributed randomly amidst empty sites ($E$), all in equal proportion. We conduct Monte Carlo simulations by randomly selecting a player $x$ from the population, who acquires its accumulated payoff $\Pi_{x}$ by playing the game with its four nearest neighbors. We refer to the latter as the $R_1$ interaction range, but we also consider $n-$level interactions ($R_n$) that encompass all players whose distance in taxicab metric is smaller than $n$ lattice constants from the focal player. Practically, it means that for $R_2$ a focal player interacts with $12$ players and for $R_3$ it interacts with $24$ players. As the key parameter we introduce the survival threshold $Q$, which characterizes the necessary demand of the environment. If $\Pi_{x} \geq Q$ player $x$ is allowed to place an offspring on a randomly chosen empty site within its interaction range (in the absence of an empty site player $x$ survives but does not place an offspring). If $\Pi_{x}<Q$ player $x$ dies out, leaving behind an empty site. Each full Monte Carlo step gives a chance to every player to either reproduce (survive in the absence of space) or die once on average.

Three qualitatively different scenarios exist depending on $Q$. If $Q>\Pi_{x}$ even for the most individually favorable strategy configurations (for example a defector being surrounded solely by cooperators in the PD game), the entire population dies out. If $Q$ is very low, both strategies are viable and essential spread randomly as determined by the toss of a coin. For intermediate values of $Q$, however, fascinating evolutionary outcomes emerge. Figure~\ref{wire} summarizes the possibilities on the $T-S$ plane for two different values of $Q$. For $Q=2$ (left) cooperators dominate across the PD and SH quadrant since the threshold is high enough to support only cooperative behavior. Defectors are able to prevail mostly in the SD quadrant, where they benefit from the game parametrization that awards cumulatively highest payoffs to mixed $C+D$ pairs. If the threshold is lowered to $Q=1.5$, defectors are able to survive in the presence of cooperators even if $S<0$. Counterintuitively, increasing the value of $S$ can support defectors before the transition to the SD quadrant. This is akin to the phenomenon observed in games of cyclic dominance, where the direct support of prey will often strengthen predators \cite{frean_prsb01, mobilia_jtb10}. In our case, if the threshold is not too high, cooperators (prey) can survive even if they are exploited by defectors (predators). Increasing $S$ supports cooperators, but if they are surrounded by defectors the latter are supported even more since they can continue with their exploitation. Thus, relatively low $Q$ can sustain a mixed state even if the overall productivity of the population is marginal. It is important to note that these results are in stark contrast with the outcome obtained if the evolutionary process is governed by strategy imitation or other proportional rules where the more successful individuals produce offspring with higher probability (see e.g. \cite{roca_plr09}).

\begin{figure}
\begin{center}
\includegraphics[width = 8.5cm]{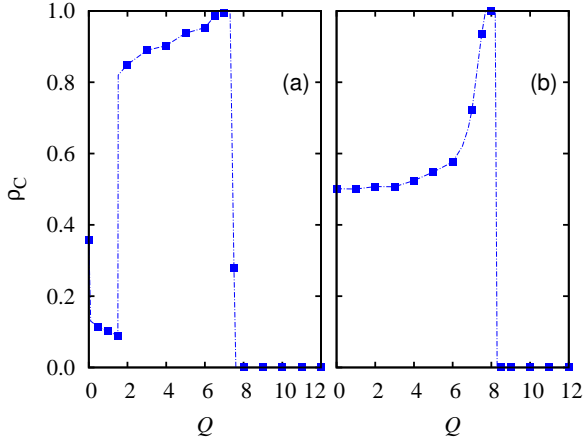}
\caption{\label{threshold}Stationary frequency of cooperators $\rho_C$ in dependence on the threshold value $Q$ for two representative $T-S$ pairs. Panel~(a) shows the result for the prisoner's dilemma game obtained by using $T=1.5$ and $S=-0.5$, while panel~(b) shows the result for the snowdrift game obtained by using $T=1.5$ and $S=0.5$. Regardless of game parametrization, there exists an intermediate interval of $Q$ values within which cooperation thrives best. We have used the $R_2$ interaction range on a square lattice with linear size $L=3200$ and random initial conditions. The displayed results are averages over $100$ independent runs.}
\end{center}
\end{figure}

To illustrate the dependence on $Q$ explicitly, we show in Fig.~\ref{threshold} the stationary frequency of cooperators $\rho_C$ for the prisoner's dilemma and the snowdrift game at $R_2$. The wider interaction range as used in Fig.~\ref{wire} allows us to observe a broader range of the $Q$ dependence, because the maximally attainable payoff of the players is accordingly larger. Regardless of this, however, the main features of the outlined threshold dependence are robust and obviously independent of the applied interaction range. It can be observed that there exists an optimal intermediate interval of $Q$ values, for which the level of cooperation in the population is maximal. Interestingly, and in agreement with our previous observations related to the results presented in Fig.~\ref{wire}, harsh external conditions that are constituted by the prisoner's dilemma confer a stronger evolutionary advantage to the cooperators.

\begin{figure}
\begin{center}
\includegraphics[width = 8.5cm]{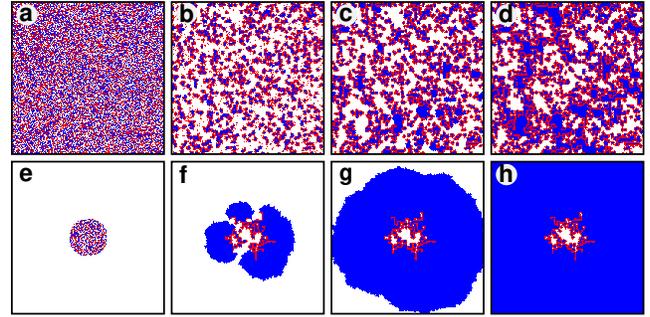}
\caption{\label{initial}Evolution of cooperators (blue) and defectors (red) from an initially fully (a-d) and partially (e-h) populated square lattice with $R_1$. Although the distribution of strategies and empty sites (white) is the same within the populated area, the final outcome is significantly different. Defectors need the presence of cooperators to survive (b,d), and they stifle proliferation of cooperative behavior in the absence of free expansion ranges (d). If the latter exist (e), cooperators spread successfully (f,g), and their compact domains are not penetrable by defectors (h). Parameter values are $L=200, T=1.5, S=-0.2$, and $Q=1.5$. Snapshots were taken after 50, 100, and 200 $MCS$.}
\end{center}
\end{figure}

Snapshots presented in Fig.~\ref{initial} reveal fundamental differences in the way cooperators and defectors spread if compared to imitation-based dynamics. In the latter case, cooperators form compact clusters to protect themselves against the exploitation by defectors (see e.g. \cite{nowak_n92b}). Binary birth-death dynamics also introduces clusters of cooperators (c,d), which allow defectors who manage to cling onto one of the interfaces to earn sufficiently high payoffs. Those defectors that fail to do so die out. This fact is illustrated clearly in the bottom row, where an initial state with ample expansion range (e) allows cooperators to free themselves from defectors (f) and expand unboundedly in a self-organized way (g,h). Defectors remain frozen within their initial radius of existence and cannot spread further. The random initial state depicted in (a), which was also used for producing Fig.~\ref{wire}, is thus not necessarily the most compatible with birth-death dynamics, as it may fail to reveal the entire potential of cooperative behavior. If a smaller portion of the lattice is used like in (e), then the final outcome will in fact always be a cooperator-dominant state, regardless of the values of $Q$ (assuming of course $Q$ is within reach of a cooperative domain), $T$ and $S$. Here a destructive strategy like defection cannot expand towards the empty space, but can only exploit other cooperators locally to ensure its own mere existence.

\begin{figure}
\begin{center}
\includegraphics[width = 8.5cm]{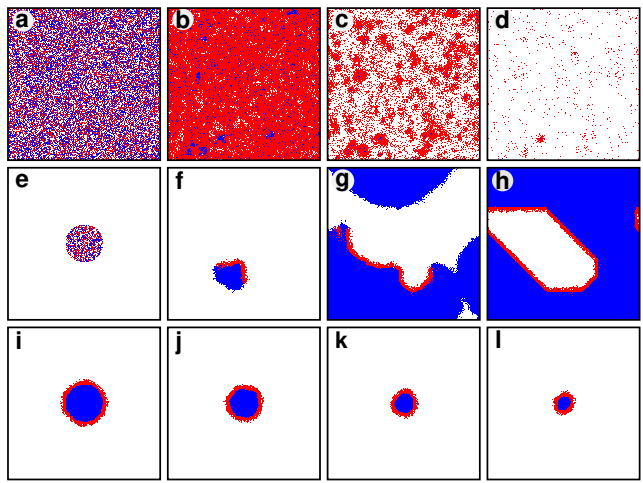}
\caption{\label{range}Increasing the interaction range from $R_1$ (see Fig.~\ref{initial}) to $R_4$ (four times the elementary nearest-neighbor range) increases the vulnerability of cooperators. In the absence of free expansion ranges (a) defective behavior thrives (b), but as soon as cooperators vanish defectors alone are unable to reach the survival threshold $Q$ and die out (c,d). If free expansion ranges do exist (e), chances are that some cooperators can evade the exploitation and spread in a given direction (f). However, defectors that succeed in maintaining contact to cooperators form a persistent front, the width of which cannot exceed the interaction range (g,h). Evolutionary stability is determined by the curvature of the invasion front, which leads to symmetry breaking in favor of defection if it is positive (i-l) (see Fig.~\ref{symmetry}). Parameter values are $L=200, T=2, S=-1$, and $Q=29$ (the higher threshold is possible due to the larger interaction range).}
\end{center}
\end{figure}

Further supporting these conclusions are snapshots presented in Fig.~\ref{range}, where the larger $R_4$ interaction range allows defectors to exploit cooperators more efficiently. In the absence of free expansion ranges (a) defectors initially thrive (b), but subsequently fall victim to the absence of cooperators and die out (c,d). This is a vivid demonstration of an actual tragedy of the commons \cite{hardin_g_s68}. Initial conditions in (e) yield a very different evolutionary outcome. Although the majority of the population within the circle dies out, some cooperators at the edge nevertheless succeed in forming a homogeneous domain and reaching $Q$ (f). They spread towards the empty space, but defectors form a persistent front that never vanishes (g,h). It turns out that the curvature of this front is crucial for cooperators to survive. If the curvature is positive (i), symmetry breaking favors defection and the whole population is again destined to die out --- as soon as cooperators vanish, defectors are soon to follow, as illustrated in (c,d).

\begin{figure}
\begin{center}
\includegraphics[width = 8.5cm]{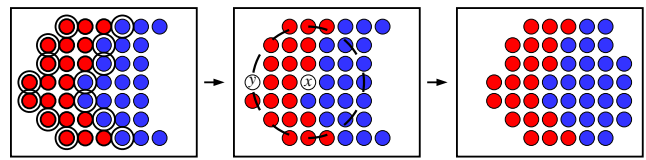}
\caption{\label{symmetry}Schematic explanation of symmetry breaking of domain wall movement that preserves the constant width of invasion fronts of defectors depicted in Fig.~\ref{range}. Cooperators (blue) that are unable to reach $Q$ and defectors (red) that are unable to reach $Q$ if a cooperator dies out are encircled (left). When cooperator $x$ dies, the empty site can be populated by an offspring from any cooperator that is within the interaction range (dashed circle in the middle panel). But since the interface has a positive curvature from the point of view of cooperators, there are even more defectors to place an offspring on $x$. Due to the symmetry breaking it is more likely that the site $x$ will be populated by a defector. As soon as this happens, however, a defector at the edge of the interface will die out (site $y$). And in this way, all the encircled cooperators in the left panel will be replaced by a defector, while all the encircled defectors will die out. At the same time, cooperators will spread further towards the empty space (right). Thus, the domains move and are stable, but the width of the defective front remains constant.}
\end{center}
\end{figure}

Details of the symmetry breaking are explained in Fig.~\ref{symmetry}. Accordingly, from the viewpoint of cooperators the interface is stable if it is either straight or has negative curvature. The defector's point of view is slightly different because $T>R$ breaks the symmetry, and a defector can thus reach $Q$ even if the curvature is positive. As panels (g,h) of Fig.~\ref{range} demonstrate, straight and negative defective fronts indeed fail to upset cooperators, while a belt of defectors can easily suffocate a cooperative domain [compare panels (i) and (l) in Fig.~\ref{range}]. We note that the curvature-dependent propagation of the $C-D$ interface is more robust at larger interaction range ($R_3$ or $R_4$) because the width of the defective front at $R_1$ or $R_2$ is very narrow and may thus break apart easily.

\begin{figure}
\begin{center}
\includegraphics[width = 8.5cm]{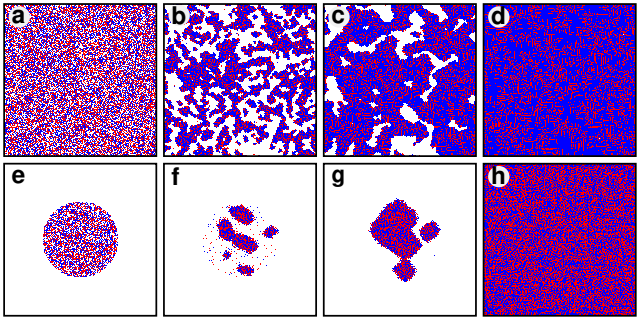}
\caption{\label{snow}If the prisoner's dilemma is replaced by the snowdrift game, cooperators (blue) and defectors (red) have the potential to spread independently. But since the mixed $C+D$ pairing yields the highest cumulative payoff above the $T+S=2$ line, a sufficiently high threshold value $Q$ can evoke the optimal solution. As panels (a-d) illustrate, the role separation emerges spontaneously from a random initial state, and although the birth-death rule acts locally, the stationary state is globally optimal. This emergence is clearer still if the $R_1$ interaction range (a-d) is replaced by the $R_4$ interaction range (e-g), and if initially the population is bounded to a small fraction of the whole lattice (e). While a random mixture of strategies fails to preclude extinction (f), spontaneously emerging compact $C+D$ clusters grow unbounded. This is a self-organizing growth. Defectors are unable to survive alone --- they need cooperators to utilize on the high value of $T$ --- but the relatively high value of $S$ sustains cooperators as well. Parameter values are $L=200, T=2$, $S=1$, and $Q=3$ (a-d) $29$ (e-h). The systems reached the final state after 500 $MCS$.
}
\end{center}
\end{figure}

Having revealed the mechanisms that may sustain the dominance of cooperators even in the most adverse social dilemma ($T=2$, $S=-1$), it remains of interest to study the evolutionary dynamics in the SD quadrant. Snapshots presented in Fig.~\ref{snow} demonstrate self-organized growth towards the globally optimal mixed $C+D$ state. It is worth noting that a similar ``role-separating'' distribution of strategies could be the result of evolution when other-regarding preferences are introduced among players \cite{szabo_jtb12}. While a random mixture of cooperators and defectors is unable to survive (b), especially not at $R_4$ (f), once the role-separation sets in, the mixed clusters with proper distribution of strategies expand until the whole population is occupied (c,d and g,h). Although the applied binary birth-death dynamics affects exclusively individual players, a sufficiently high threshold value $Q$ spontaneously selects the global optimum through self-organization.

Summarizing, we have shown that replacing imitation with a binary birth-death rule in spatial evolutionary games creates a new class of solutions of social dilemmas. If free expansion ranges are paired with limited exploitation possibilities, cooperative behavior dominates the prisoner's dilemma and the stag-hunt game by means of unbounded self-organized expansion that sets in as soon as cooperators find a niche to expand. If defectors are given the opportunity to exploit cooperators more effectively through the application of larger interaction ranges, cooperative behavior may still thrive, although it relies on a special type of symmetry breaking that determines the direction of invasion based on the curvature of the interface that separates the two competing strategies. Counterintuitively, higher levels of cooperation are observed in the prisoner's dilemma than in the snowdrift quadrant, since in the latter case self-organized growth favors mixed $C+D$ domains with proper role separation. Regardless of the governing social dilemma, and despite of the fact that the birth-death dynamics acts locally, the stationary state is always globally optimal in that the payoff of the entire population is maximal. We note that all the presented results are robust to variations of the interaction network, and can be observed also in off-lattice simulations \cite{antonioni_prep}, thus indicating a high degree of universality. Yet the studied model also invites a more sophisticated introduction of separate thresholds for births and deaths to be studied in the future, although we expect that this will not influence the fundamental positive role of the reported self-organized growth. Another promising avenue to explore in the future could be the combination of imitation and binary birth-death rules, which might give rise to further fascinating evolutionary outcomes that are driven by pattern formation.

Our results also highlight the importance of initial conditions, in particular the fact that a random initial state without obvious chances of expansion may fail to reveal all the benefits of cooperative behavior. This may be relevant in experimental setups. Recent experiments with microbial metapopulations indeed support the conclusion that range expansion may promote cooperation \cite{manoshi_pnas13}. Although special initial conditions with ample empty space may appear somewhat artificial in the realm of mathematical modeling, they do mimic rather accurately the conditions in a Petri dish \cite{elena_nrg03,van-dyken_cb13}, and together with birth-death dynamics they appear to hold the key for understanding the success of cooperation from an entirely different perspective.

\acknowledgments
This research was supported by the Hungarian National Research Fund (Grant K-101490), the Swiss National Scientific Foundation (Grant 200020-143224) and the Slovenian Research Agency (Grant J1-4055).

\end{document}